\newcommand{\beq}{\begin{equation}}
\newcommand{\eeq}{\end{equation}}

\newcommand{\bear}{\begin{eqnarray}}
\newcommand{\eear}{\end{eqnarray}}

\RequirePackage[switch]{lineno}

\documentclass[review]{elsart5p}

\usepackage{graphicx}
\usepackage{lscape}
\usepackage{enumerate}
\usepackage{pifont}
\usepackage{esvect}
\usepackage{amssymb}
\usepackage{bbding}
\usepackage{rotating}
\usepackage{footnote}
\usepackage{tablefootnote}
\usepackage[flushleft]{threeparttable}
\usepackage{amsmath}
\usepackage{color}
\usepackage[dvipsnames]{xcolor}
\usepackage{amsfonts}
\usepackage{dsfont}

\usepackage{url}
\usepackage{tabularx}

\usepackage{multicol}
\usepackage{float}
\usepackage{booktabs}
\usepackage{xspace}
\usepackage{pbox}
\usepackage{verbatim}

\usepackage{color}\definecolor{darkblue}{rgb}{0,0,1}

\makesavenoteenv{tabular}
\makesavenoteenv{table}

\begin{document}

\begin{frontmatter}

\title{Time and band-resolved scintillation in time projection chambers based on gaseous xenon}

\bigskip
\author[1]{S. Leardini}, \author[2]{E. S\'anchez Garc\'ia$^{1,}$}\thanks{
Currently at Max-Planck-Institut f\"ur Kernphysik, Saupfercheckweg 1, 69117 Heidelberg, Germany}, \author[1]{P. Amedo}, \author[1]{A. Saa-Hern\'andez}, \author[1]{D. Gonz\'alez-D\'iaz\corauthref{aut1}\ead{Diego.Gonzalez.Diaz@usc.es}}, \author[2]{R. Santorelli}, \author[1]{D. J. Fern\'andez-Posada}, \author[1]{D. Gonz\'alez}

\corauth[aut1]{Corresponding author.}
\address[1]{Instituto Galego de F\'isica de Altas Enerx\' ias (IGFAE), Universidade de Santiago de Compostela, Spain}
\address[2]{Centro de Investigaciones Energéticas, Medioambientales y Tecnológicas (CIEMAT), Madrid, Spain}

\begin{abstract}
We present a systematic study of the time and band-resolved scintillation in xenon-based time projection chambers (TPCs), performed simultaneously for the primary (S1) and secondary (S2) components in a small, purity-controlled, setup. We explore a range of conditions of general academic interest, focusing on those of relevance to contemporary TPCs: pressure range $1$-$10$~bar, pressure-reduced electric fields of $0$-$100$~V/cm/bar in the drift region (S1) and up to the proportional scintillation regime in the multiplication region (S2), and wavelength bands 145-250/250-400/400-600~nm, for both $\alpha$ and $\beta$ particles. Attention is paid to the possibility of non-conventional scintillation mechanisms such as the $3^{rd}$ continuum emission, recombination light from $\beta$-electrons at high pressure (for S1), emission from high-lying excited states and neutral bremsstrahlung (for S2). Time constants and, specially, scintillation yields have been obtained as a function of electric field and pressure, the latter aided by Geant4 simulations.
\end{abstract}

\begin{keyword}
Time Projection Chambers \sep TPCs \sep drift chambers \sep imaging chambers

\PACS 29.40 \sep Cs
\end{keyword}
\end{frontmatter}
\section{Introduction} \label{intro}
The main characteristics of xenon scintillation (and of noble gases at large) are nowadays well established thanks primarily to measurements performed with selective photo-excitation \cite{photoex1, photoex2, photoex3, photoex4, photoex6, photoex7, photoex8, photoex9}, synchrotron radiation \cite{synch}, electric discharges \cite{Plasma1, Plasma2, Disc1}, and bunched beams of x-rays \cite{xrays}, electrons \cite{Koehler, Langhoff_e}, protons \cite{Hurst} and heavy ions \cite{heavy1, Wiese}. Time or band-resolved measurements in conditions of interest to radiation technology are, however, much more scarce: among them, studies under $\alpha$-particles \cite{suzu1, suzu2, Millet} and, more recently, $\beta$-electrons \cite{japacryo} can be highlighted.
  
The dominant scintillation mechanism above some 100's mbar of xenon is the so-called $2^{nd}$ continuum emission, with a spectrum centered at around 172~nm and with an approximately Gaussian width of 10~nm (e.g., \cite{Koehler}). If attending to uncertainties reported in the careful and independent investigations performed most recently in \cite{Japa1, Japa2, Michel}, the average energy needed for an $\alpha$-particle to produce a photon around this band, for electric fields high enough to suppress charge recombination, is nowadays known to a precision of around one eV in the region 1-10~bar: $W_{sc}=35.0\pm 1.2~$eV. Although uncertainties in $W_{sc}$ for x-ray and electron excitation have been historically reported to be considerably larger, in the range $60$-$110$~eV \cite{FernandesW, RennerW, ParsonsW, doCarmoW}, recent studies point to values compatible with the ones measured for $\alpha$ particles \cite{HenriquesUpcoming}. Furthermore, state of the art calculations of the seed states stemming from x-ray and MeV-electron excitation \cite{DEGRAD}, once coupled to a microscopic description of the atomic/excimer cascade, can reproduce the main features of the scintillation spectrum of weakly quenched xenon mixtures \cite{XeSim}, providing $W_{sc}=39\pm 1$~eV in the pressure range 1-10~bar. Calculations provide time and band-resolved spectra in a range of pressures from 0.1 to 10~bar and wavelengths from 145~nm to 1300~nm, for \%-levels of common molecular additives. The same framework enables the computation of the secondary scintillation produced by low-energy ionization electrons accelerated in electric fields (a calculation undertaken for pure noble gases earlier in \cite{Oliveira}). As the gaseous transport of those ionization electrons in electric and magnetic fields is, too, firmly established nowadays \cite{Magboltz, Garfield++, Pyboltz}, the experimental situation would seem satisfactory for the conditions of interest to xenon-based gaseous detector technology.

With the increased understanding of the transport and scintillation characteristics of weakly-quenched xenon mixtures, and in light of the necessity of a better understanding of existing devices and exploration of their technological limits (for a review, see e.g. \cite{DiegoTPC}), attention has been brought recently to a number of subdominant and intriguing scintillation mechanisms, hitherto neglected. The `$3^{rd}$ continuum' emission, for instance, has been proposed in \cite{Edgar} as a means to perform electron/nucleus separation in argon TPCs, with potential use in direct dark matter searches. Along the same line, for TPCs that are strongly reliant on the electron scintillation such as NEXT \cite{NEXT}, the absolute scintillation yield and spectral content is crucial, as it determines how far from the cathode a low-energy krypton event can be reconstructed during energy calibrations \cite{NEXT_Kr}. In xenon, where $3^{rd}$ continuum emission in the 250-400~nm band is well established since the works of Henck \cite{Henck} and, specially, Millet \cite{Millet}, information on the scintillation yields and their dependence with pressure and particle type have not yet been published, to the best of our knowledge. In practice, and given the higher quantum efficiency and better optical properties involved in their detection (less absorption to gas impurities and better reflectivity at surfaces), it can be expected that the $3^{rd}$ continuum emission represents a substantial fraction of the primary scintillation in noble-gas TPCs, dominating even over the fast component in some conditions (e.g. in argon in the range 1-5~bar \cite{Edgar}). On another front, the NEXT collaboration recently discussed the possible presence of charge recombination for MeV-electron tracks at pressures as high as 10~bar \cite{NEXTreco} and, despite measurements for X-rays have excluded recombination effects up to 10~bar in \cite{Balan}, the impact of the phenomenon for calorimetry in $\beta\beta0\nu$ experiments is important enough that it requires further elucidation. Another not-understood issue is the observed violation of pressure-reduced scalings of the proportional scintillation yields (electroluminescence, EL) in xenon \cite{Pviolation}, greatly exceeding the expectation from the departure of the ideal gas behaviour. At typical EL fields around 1-2~kV/cm/bar, the deviation can represent up to a 70\% increase in the scintillation yields when going from 1 to 10~bar. Last, the observation of neutral bremsstrahlung scintillation in argon TPCs \cite{nBrAr} has opened the path to similar explorations in xenon TPCs, where the phenomenon was unambiguously demonstrated to great precision in the secondary scintillation signal \cite{nBrXe1, nBrXe2}. Experimental information about its spectral content is not available as of today, though. 

Motivated by the aforementioned developments, we assembled a purity-controlled mini-TPC dedicated to the systematic study of the time and band-resolved features of primary (S1) and secondary (S2) scintillation in xenon. The layout of the paper is straightforward: section \ref{SectionSetup} describes in detail the setup, analysis and simulation procedures performed in order to extract the absolutely normalized time spectra for each particle type, spectral band, pressure and field, section \ref{S1} presents S1 data and section \ref{S2} presents S2 data.

\section{Experimental procedure \label{SectionSetup}}

\subsection{Description of the setup}
\begin{figure}[h!!!]
    \centering
    \includegraphics[scale=0.75]{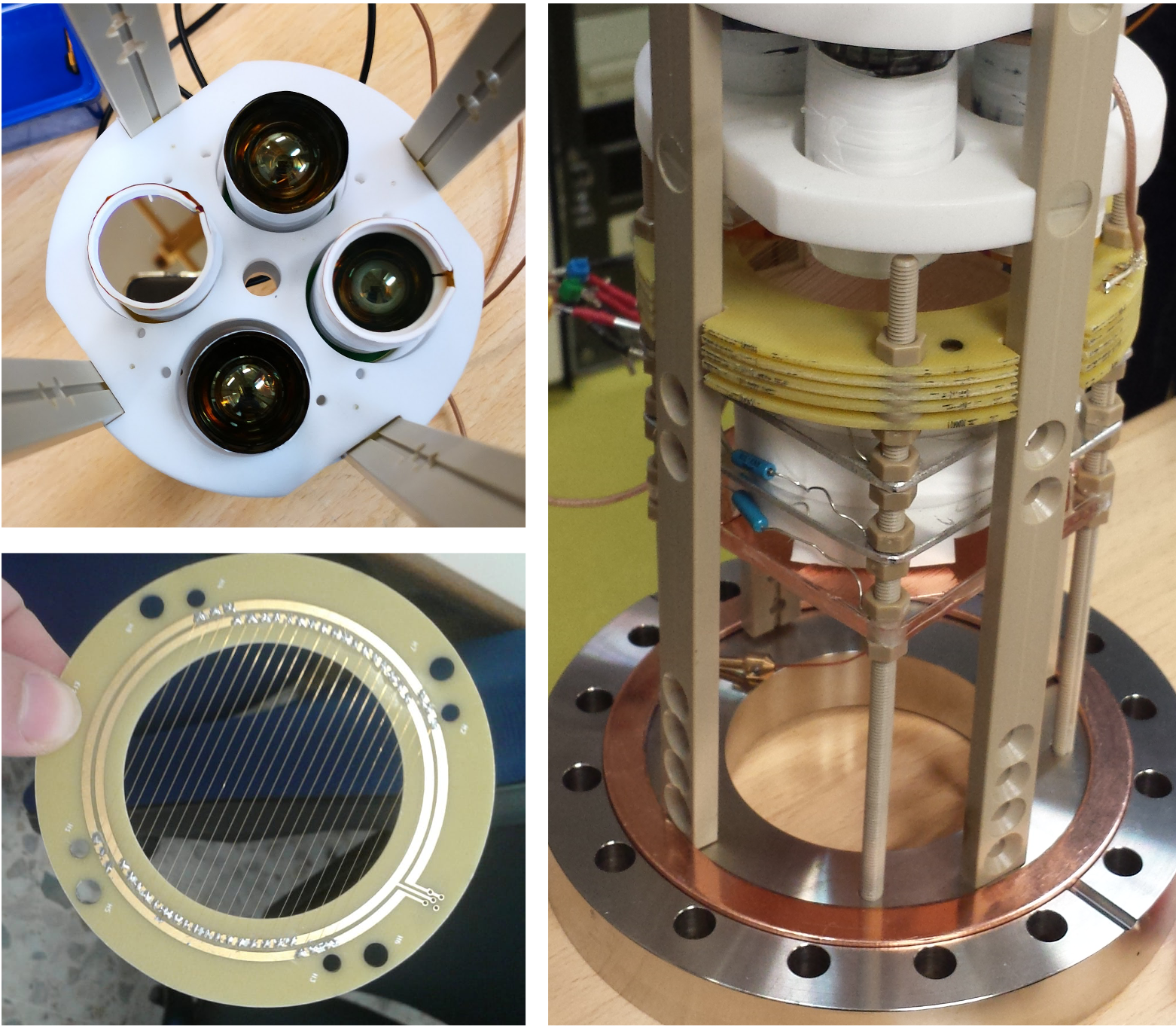}
    \caption{Some key elements of the setup used in the present measurements. Left-top: pressure-resistant PMs (R7378 from Hamamatsu) with band-pass filter (positioned left-right in the frame) and without (top-bottom). Left-bottom: a cathode from one of the multi-wire proportional chambers. Right: the setup once assembled and prior to mounting inside the vessel.}
\label{fig:setup}
\end{figure}

The pressure/vacuum system employed in the present measurements has been described elsewhere \cite{FATGEMs}: it features a high-pressure compressor (10 normal liters per minute at 12~bar) to force gas circulation through a cold getter (SAES MC190-903FV), and sampled with a leak valve into a residual gas analyzer for purity assessment (SRS-RGA~200). Xenon is cryorecovered periodically with liquid nitrogen so that volatile impurities oblivious to the cold getter (mostly N$_2$), and that accumulate due to system outgassing, can be pumped away. The experimental setup consists of a mini-TPC (4.5~cm drift, 6.5~cm diameter) whose anode region is instrumented with a multi-wire proportional chamber (MWPC), see Fig. \ref{fig:setup}. 

\begin{figure}[h!!!]
    \centering
    \includegraphics[scale=0.35]{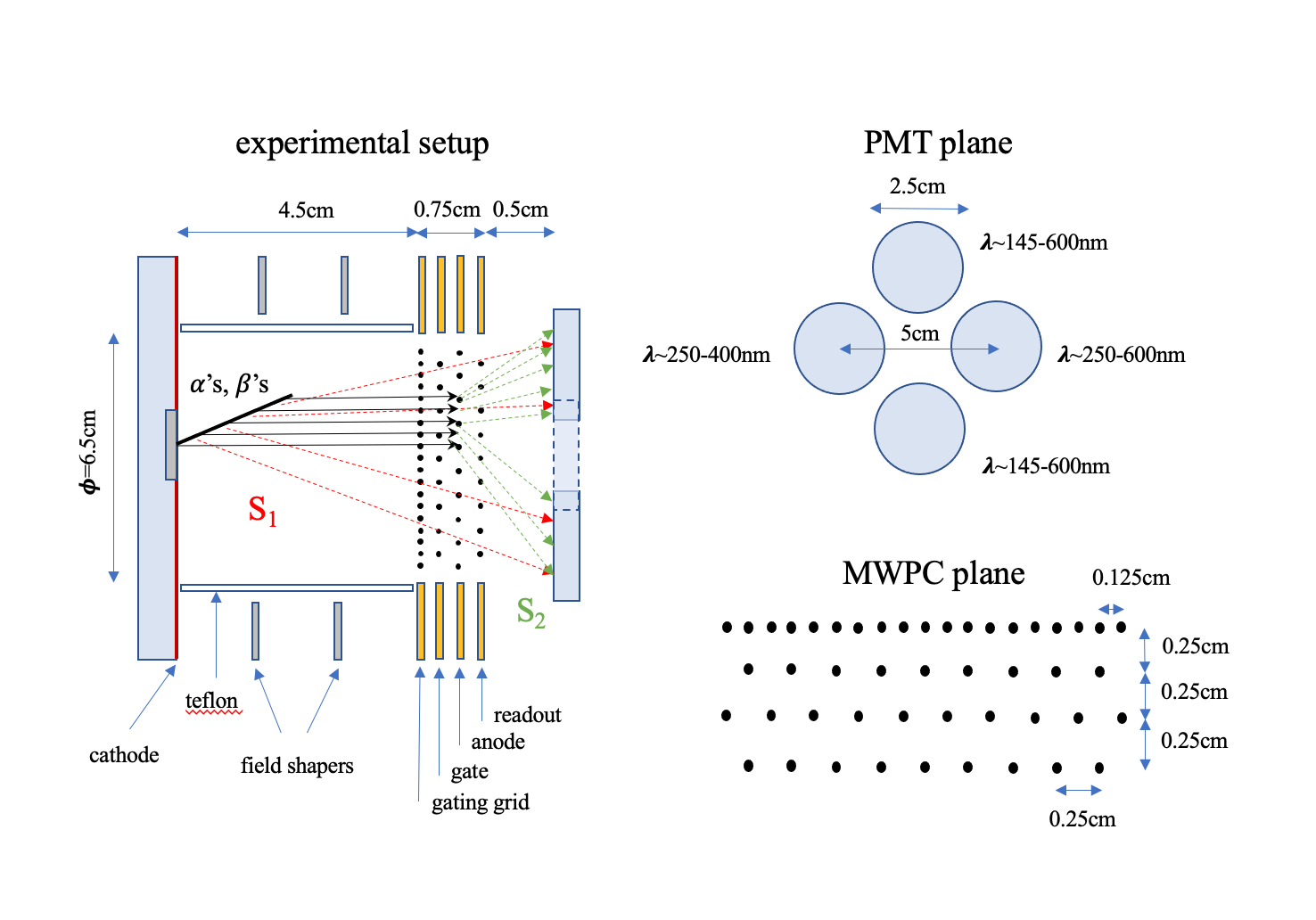}
    \caption{Left: sketch of the experimental setup, indicating the main elements and distances, as described in text. Right-top: arrangement of various PM+filter assemblies. Right-bottom: wire plane layout.}
\label{fig:sketch}
\end{figure}

\begin{figure}[h!!!]
    \centering
    \includegraphics[scale=0.5]{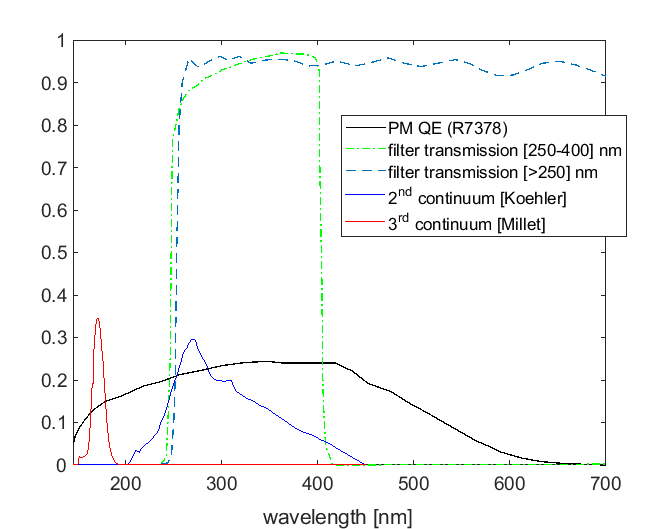}
    \caption{Wavelength response of the filters and PM used in this work. The $2^{nd}$ continuum spectra measured by Koehler \cite{Koehler} in an electron beam (red) as well as the $3^{rd}$ continuum measured by Millet \cite{Millet} at 1~bar (blue) are also shown, arbitrarily normalized.}
\label{lambda_dep}
\end{figure}

As the setup is aimed at optical readout, both MWPC-cathodes are based on wires to ensure good optical transparency towards the photosensor plane. Although immaterial for present measurements, the amplification structure has a gating grid too, that has been set at all times to the voltage corresponding to the electrostatic potential expected for an uniform field in the drift region (`open'). All four planes' wires are made of a 99.95\% purity tungsten core, plated with 99.99\% purity gold. Anode wires have a diameter of 20 $\pm$ 2 $\mu$m, and the gold plating accounts for 3-5\% of their total mass. For the remaining planes, 80~$\pm$~8~$\mu$m diameter wires with a gold coating thickness of 0.5~$\mu$m were used. Anode and cathode wires have a pitch of 2.5~mm (shifted by half a pitch for each alternating plane), while the gating grid has a pitch of 1.25~mm (a sketch of the setup is shown in Fig. \ref{fig:sketch} bottom-right). MWPCs have been fabricated at IGFAE labs based on custom-designed printed circuit boards (PCB) of about 6~cm aperture (Fig. \ref{fig:setup} bottom-left). Wires were aligned with the help of 200~$\mu$m-wide slits drilled on an aluminum jig, weights applied to them (corresponding to tensions around 50 grams for the anode, and 100 grams for the cathode) and finally soldered to PCB pads. Analysis of the MWPC photographs confirmed that the dispersion of the wires position was well within the target value of 200~$\mu$m. 
Metallic disks with deposits of $^{241}$Am (for $\alpha$ runs) or $^{90}$Sr (for $\beta$ runs) were housed alternatively in a circular groove at the cathode center, making sure to leave them flush with the cathode surface, and thus ensuring a minimal distortion of the electric field. Field homogeneity was enhanced through two circular aluminum frames placed along the drift dimension (`shapers'), separated by 1.5~cm and connected with resistors. Their inner radius (facing the TPC active volume) was layered with a 1~mm-thick reflector made of expanded teflon. Despite the modest 20\%-level reflectivity anticipated \cite{TefRef}, its use facilitates the simulation of the setup and interpretation of the results, avoiding parasitic reflections from the shapers or chamber vessel. Behind the anode, four pressure-resistant photomultipliers (PMs), model R7378 from Hamamatsu, were assembled in a teflon frame and placed at close distance (Fig. \ref{fig:setup} top-left). This PM model has a maximum quantum efficiency of 24\% at 400~nm, dropping by about a factor $\times 10$ outside the 145-600~nm range that is the sensitivity range assumed hereafter. Two filters were used, aimed at identifying emissions in the NUV and visible regions (250-400~nm and $>$250~nm). The response of both filters and PMs is presented in Fig. \ref{lambda_dep}, together with $2^{nd}$ and $3^{rd}$ continuum spectra from \cite{Koehler} and \cite{Millet}. 

Since the third continuum measured earlier for $\alpha$-particles in \cite{Millet, Henck} peaks at 270~nm, with a few 10's of nm -wide blue-wing and a red-wing extending up to wavelengths slightly above 400~nm, the present choice of filters is convenient. Little sensitivity above 600~nm can be expected in these measurements though, due to the lack of PM efficiency. In the following we labelled these PM+filter assemblies by their response as `145-600', `250-400', `250-600'. Data acquisition was performed with a CAEN DT5725 board having 14~bits, 125~MHz bandwidth and a sampling time of 4~ns. Experience shows that, with 10's of thousands events, this bandwidth and binning is sufficient to reconstruct exponential time constants down to at least 4~ns (e.g., \cite{Edgar}), as needed for present studies.
The acquisition window was set to 5~$\mu$s for S1 data (in order to account for the charge recombination tail at high pressures) and to 12~$\mu$s for S2 data (in order to fully include the spatial extent of the ionization trail for all conditions studied). Combined S1-S2 runs were performed to evaluate the electron drift velocity and consistency with the separate S1, S2 analyses. The board was triggered by PM1[145-600] at a voltage threshold corresponding to around 2-4~phe. A charge threshold at 6~phe was then applied by software.

A photograph of the experimental setup prior to its assembly in the high pressure vessel is shown in Fig. \ref{fig:setup}-right, and a sketch of the system is given in Fig. \ref{fig:sketch}. The chamber dimensions were chosen to provide good containment of 5.5~MeV $\alpha$-tracks down to 1~bar ($R_\alpha$=2.2~cm/bar \cite{NIST}) and $0.5$~MeV-electrons at 10~bar ($R_e \sim 25~$cm/bar \cite{Accu}), at the same time fitting inside a CF100 vessel, equipped with all the necessary service ports (leak valve, pumping port, gas inlet/outlet, PM-HV feedthroughs, PM-signal feedthroughs, TPC-HV feedthroughs, vacuum gauge). All connections are based on metal-metal VCR and CF standards.

During the measurements the chamber was pumped until a vacuum of $10^{-3}$~mbar was achieved and then filled with xenon (purity-grade 5, from Nippon gases). After establishing recirculation, few minutes were sufficient for the scintillation yield to steadily increase until it reached saturation. At that point, a purity level down to $<$1~ppm of O$_2$ could be estimated from the RGA reading when sampling with a leak valve, whereas the H$_2$O and N$_2$ levels in the vacuum region were too high to allow measurements below 1000~ppms. Based on the O$_2$-levels and given that the getter removes H$_2$O, O$_2$ and other highly reactive impurities down to ppb-level after a single pass, the main contaminant in the system is expected to be N$_2$, that was purged periodically with cryorecovery/pumping cycles. In general, the triplet time constant of the second continuum stayed in the range $\tau^*=$85-98~ns during the measurements (compared to a world average of $\tau = 100.9\pm0.7$ ns cited for instance in \cite{nBrXe1}). This means, if using the quenching rates of N$_2$ in xenon from \cite{Setser} and the formulas in \cite{XeSim}, that N$_2$ concentrations can be estimated to be in the range $f_{N_2} \simeq 300$-$500$~ppms, following the standard relations:
\bear
& \frac{1}{\tau^*} & = \frac{1}{\tau} + f_{N_2} \cdot K_{N_2} \cdot P\\
& f_{N_2} & = \frac{1}{PK_{N_2}} \left( \frac{1}{\tau^*} - \frac{1}{\tau} \right)
\label{impurities_eq}
\eear
Here $K_{N_2}$ is the N$_2$ quenching rate for Xe$^*$ states at $T=20$~deg (in [bar$^{-1}$ ns$^{-1}$]). Variations in the observed electron drift velocity using S1-S2 time differences and comparison with Pyboltz simulations yielded results compatible with N$_2$ being present in the system at the level of 100's of ppm too, no other common contaminants being able to explain the observations.

These relatively high values compared to the initial purity of the bottle (10~ppm) seem to be a consequence of the chosen setup, that integrates (besides low-outgassing PEEK bars) structural plastics like teflon, acrylic and FR4 for the MPWC structures, PMs and specially their bases, within a small volume. In the absence of hot getters, N$_2$ accumulates over time until reaching equilibrium. The justification for this experiment's design is that in xenon (contrary to argon), N$_2$ is known not to cause scintillation in Xe-mixtures even for concentrations up to percent level \cite{Taka}, that is satisfied in our setup by more than one order of magnitude in all cases. Moreover, the comparison of the observed time constants for the $3^{rd}$ continuum emission with values in the literature (as well as the observed independence on purity levels) allows us to conclude that the emissions reported here are unlikely to be affected by impurities except for the small quenching observed for the 2$^{nd}$ continuum, that amounts to 10-25\% levels. In the following, those yields are corrected for the effect using the correction factor (e.g. \cite{XeSim}):
\beq
F_{corr} = \frac{\tau}{\tau^*}
\eeq

\subsection{Data analysis}

\begin{figure}[h!!!]
    \centering
    \includegraphics[scale=0.45]{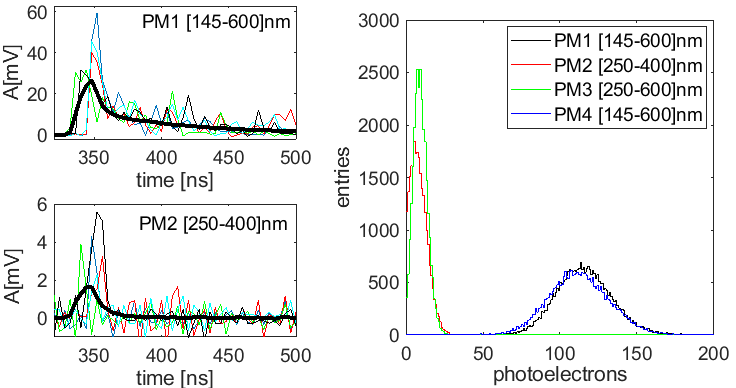}
    \caption{Left: example of primary scintillation signals acquired during an $\alpha$-run at 10~bar. Right: signal-integral distributions in the four PM, obtained in the same conditions and calibrated to photoelectrons.}
\label{fig:cuts}
\end{figure}

Data analysis was performed with Matlab scripts, including basic functions for pulse-shape analysis such as baseline corrections and waveform time-alignment to correct for offsets. Characteristic waveform parameters (rise-time, fall-time, amplitude, integrated charge) were then retrieved. The threshold used in the analysis is 6~phe, chosen to be comfortably above the levels used for triggering the DAQ card. Outlier events may be removed through soft cuts in the charge, position (from the charge barycenter of the four PMs) and firing-time cuts. As particle rates are of the order of 100's of Hz, the influence of cuts for outlier removal was minimal on the final results. We chose to just apply a $3\sigma$-cut around the peak distribution for $\alpha$-runs while, in the absence of a peak, no cut was used for $\beta$-runs. Fig. \ref{fig:cuts}-left shows an example of several waveforms taken in a typical $\alpha$-run together with their average (thick line). The distributions of charge integrals are shown on the right.

\begin{figure}[h!!!]
    \centering
    \includegraphics[scale=0.55]{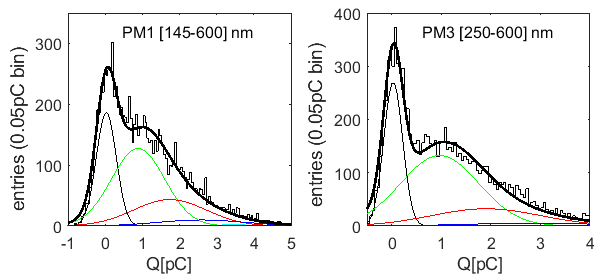}
    \caption{Signal-integral distributions of two PMs obtained in luminosity conditions corresponding to about one photon per 100~ns time window (regulated with a pulsed LED). The single-photon charge distribution (green curve) is assumed to be Gaussian and multi-photon ($n$) peaks are fitted to its $n$-convolutions, with integral values bound through Poisson statistics. The average value and uncertainty of the single-photon charge is obtained from 3-4 fits with light yields around the ones shown here.} 
\label{fig:SP_cal}
\end{figure}

\begin{figure}[h!!!]
    \centering
    \includegraphics[scale=0.6]{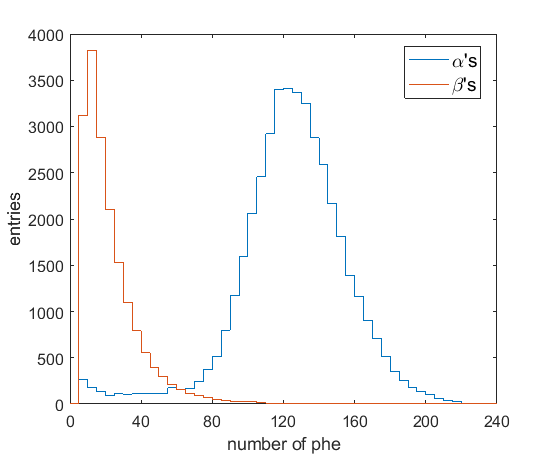}
    \caption{Spectra (in number of detected photoelectrons) for the $^{90}$Sr $\beta$-source (red) and for the $^{241}$Am $\alpha$-source (blue), obtained at 10~bar and high field conditions (above 50V/cm/bar).}
\label{fig:All_spectra}
\end{figure}

In order to convert the measured charges to photoelectrons, a blue LED was placed inside the chamber and powered with a fast pulser (model 81130A by Agilent) by means of a square signal down to 40~ns width. Data was then taken for intensity conditions estimated to be at around the single-photon level. 
A fit was performed assuming a Gaussian distribution for the PM charge-response function measured in a 100~ns window, and taking its $n$-times convolution to account for multiple ($n$) photons. The integral values of those Gaussians, including that of the pedestal, were bound through Poisson statistics for a given average number of photons. The mean and width of the PM charge-response as well as the average number of photons were left as free parameters in the fit. 
Results are shown in Fig. \ref{fig:SP_cal}. In general, the values for the centroid position and width were observed to be within $5$-$10\%$ irrespective from the LED intensity, for a Poissonian average in the range $\bar{n}_\gamma=$0.5-2. The photoelectron spectra measured in the range 250-600~nm are shown in Fig. \ref{fig:All_spectra} for $\alpha$ and $\beta$ particles at 10~bar.

With the PM signals corrected for baseline and time offsets, and calibrated, it is possible to obtain the time profiles of the emission in photoelectrons per unit time. Following the procedure sketched in \cite{Edgar}, they have been fitted to functions like:
\beq
\mathcal{F}(t) = (A_r e^{-t/\tau_s} + A_f e^{-t/\tau_f}) \circledast G(\sigma_, t_o) \label{ExpFunc}
\eeq
where $G(\sigma_, t_o)$ is the PM time-response function under $\delta$-excitation (assumed to be a Gaussian of width $\sigma$ and offset $t_0$) and the symbol $\circledast$ denotes a convolution. In the case of the unfiltered PMs (range 145-600~nm), $\tau_f$ corresponds to the effective time constant stemming from the superposition of the second continuum singlet and the third continuum (`fast'), whereas $\tau_s$ is the time constant of the second continuum triplet softly modulated by charge-recombination (`slow'). For the PMs insensitive to the second continuum signal (250-400/250-600~nm), only one time constant is needed in the fit. A more general fitting procedure should involve the formation times, but there is no sensitivity to them in present conditions and thus it destabilizes the fit. Moreover, for xenon above 2~bar the shape of the $2^{nd}$ continuum emission remains largely unchanged as shown in \cite{XeSim}.

For all four PMs a \%-level afterpulsing at a time delay around 250~ns with respect to the main signal peak was observed. The feature is reproducible with LED pulses and also observed in independent measurements performed in a dark box, confirming that it is inherent to the PM. For charge recombination studies, given the large times involved, an afterpulsing template was created by subtracting to the waveform the fit of expression \ref{ExpFunc} to the average waveform (done in the range 0-200~ns and extrapolated to higher values). For each pressure, the fit was performed in field conditions where charge recombination was negligible and the shape of the afterpulsing contribution hence estimated and subtracted for lower field values. 

In general, whenever yields are quoted, those in the range 145-250~nm refer to measurements performed in the range 145-600~nm after subtracting those for 250-600~nm (and analogously for the 400-600~nm range). For 145-600~nm, where two PMs are available, the average value is taken, their values agreeing to within less than 10\%.

\subsection{Geant4 simulations} \label{G4}

\begin{figure}[h!!!]
    \centering
    \includegraphics[scale=0.55]{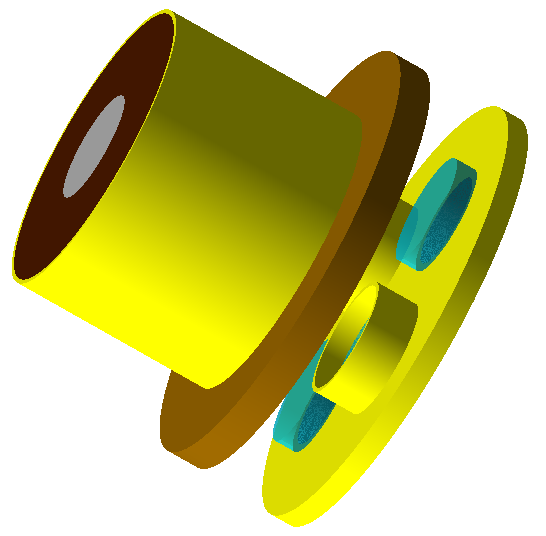}
    \caption{The experimental setup modelled in Geant4, including the copper cathode, aluminum-deposited radioactive source, cylindrical teflon reflector, MWPC frames, and the four PMs supported in a teflon frame.}
\label{fig:setup_G4}
\end{figure}

A Geant4~\cite{GEANT4} simulation model was used to quantify the light collection efficiency, which enables obtaining absolute scintillation yields. The model includes all elements in the setup (except the MWPC wires, that are included as an overall transparency factor): the copper cathode, with the source housed at its center, deposited on an aluminum disk. It also includes the cylindrical teflon reflector, the MWPC frames (FR4-based), the four PMs, filter holders and teflon frame. Each PM is modelled as a 5~mm-thick/25~mm-diameter fused silica window over a 21~mm-diameter photocathode acting as a sensitive detector. Two of the PMs are coupled to band-pass filters by means of 10~mm-long holders, as shown in Fig.~\ref{fig:setup_G4}. The optical properties have been taken from the literature, including the reflectivities at the wavelength of interest of expanded teflon~\cite{TefRef}, copper~\cite{Rcopper, Rcopper2}, aluminum~\cite{Raluminum, Raluminum2, Raluminum3}, and the refractive index of fused silica~\cite{refractiveSilica}. A table compiling the values of reflectivity for the different setup materials is presented in table~\ref{tab:reflectivities}.

\begin{table}[ht]
\centering
\caption{Reflectivity values for the different setup materials.}
\begin{tabular}{p{0.33\linewidth}p{0.33\linewidth}p{0.33\linewidth}} \toprule
    $\lambda$ & 170 nm & 270 nm \\ \midrule
    $R_{Al}$  & 0.70 - 0.75 & 0.80 - 0.90 \\
    $R_{Cu}$  & 0.00 - 0.05 & 0.20 - 0.35 \\
    $R_{teflon}$& 0.16 - 0.21 & 0.22 - 0.29 \\ \bottomrule
\label{tab:reflectivities}
\end{tabular}
\end{table}

\begin{figure}[h!!!]
    \centering
    \includegraphics[scale=0.33]{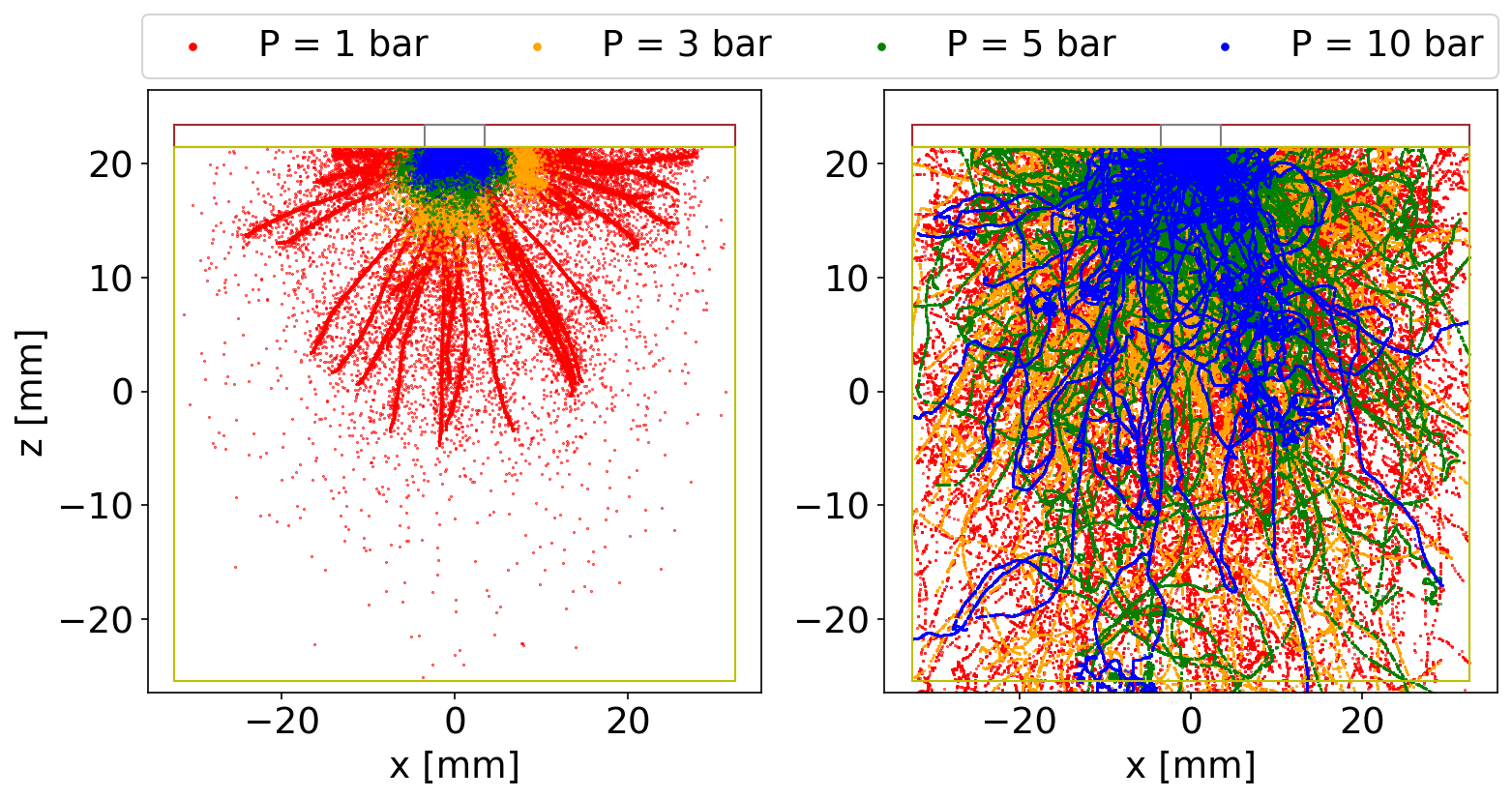}
    \caption{Simulated tracks obtained at different pressures for a cross section of the setup, for $\alpha$ (left) and $\beta$ (right) tracks emitted from the source surface (upper rectangle).}
\label{fig:sources_G4}
\end{figure}

\begin{figure}[h!!!]
    \centering
    \includegraphics[scale=0.52]{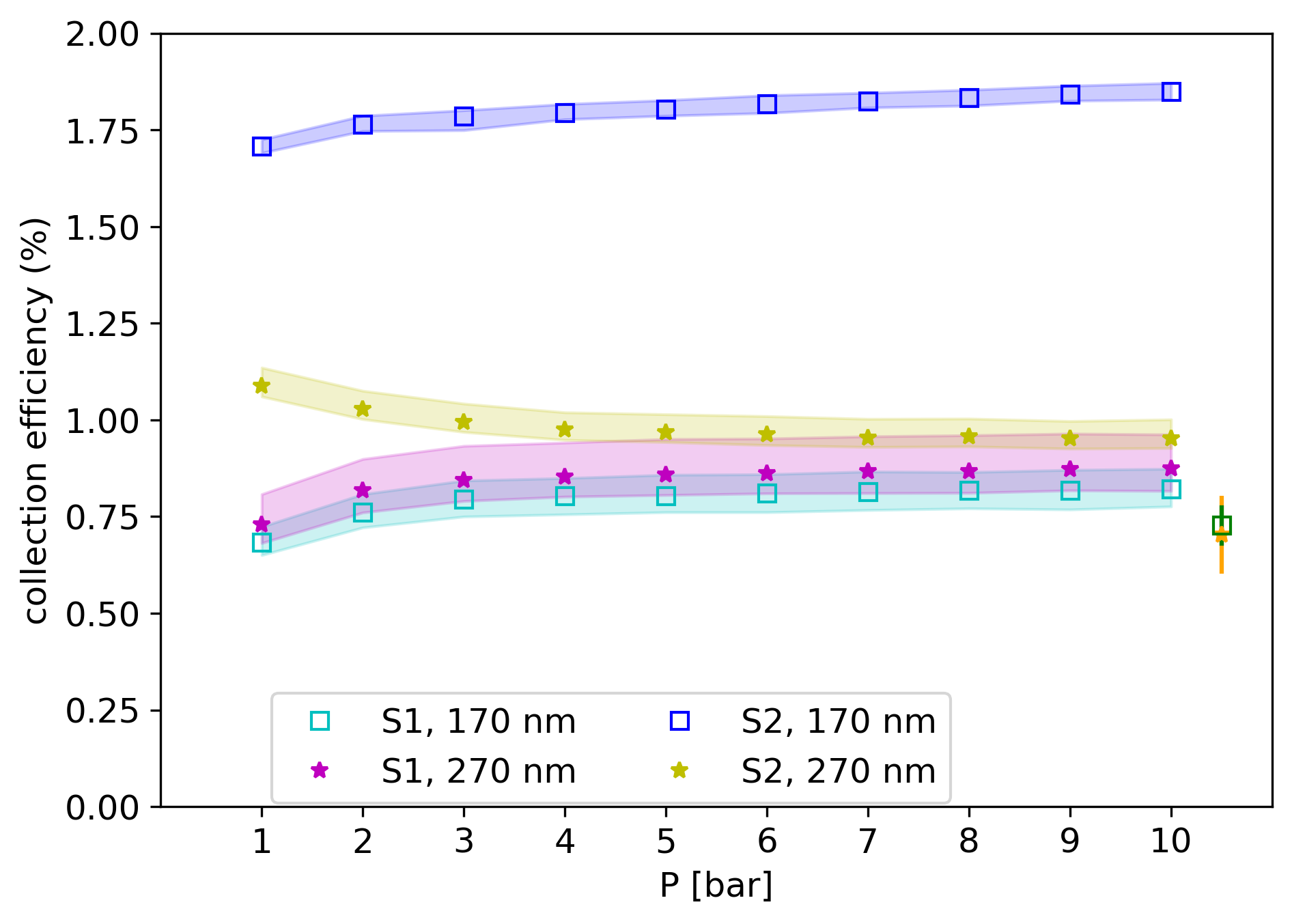}
    \caption{Simulated light collection efficiencies per PM obtained in the experimental setup for $\alpha$-tracks, as a function of pressure. For the S1 and S2 components, photons at 170~nm and 270~nm (the main peaks of the $2^{nd}$ and $3^{rd}$ continuum emission) have been employed. Error bands account for the range of parameters values.
    }
\label{fig:collection_efficiency_a}
\end{figure}

Light collection efficiency was simulated both for the S1 and S2 components at 170~nm and 270~nm, the main peaks of the $2^{nd}$ and $3^{rd}$ continuum emission. The overall dependence with wavelength was found to be small in present conditions, the dominant effect being the collimation introduced by the filter holders (see later, Fig. \ref{fig:collection_efficiency_a}). For S1, photons were generated isotropically at the interaction positions of the $\alpha$ and $\beta$ particles, using the $^{90}$Sr spectra in this latter case (Fig.~\ref{fig:sources_G4}).
For S2, photons were generated isotropically at the positions of the ionization electrons upon arrival at the MWPC anode. The spread of the ionization electrons, obtained from PyBoltz~\cite{Pyboltz} as a function of pressure and drift distance from the point of electron production to the MWPC anode, was included too.

In order to reduce the computation time due to photon tracing, and statistically represent all track topologies, their spatial distribution was estimated initially by propagating a large number of ($\alpha/\beta$)-events. The geometrical corrections were thus estimated by launching $N = 10^6$ photons from the resulting distributions. 

Calculated light collection efficiencies at 170 and 270~nm for S1 and S2 are presented as a function of pressure in Fig.~\ref{fig:collection_efficiency_a}. For $\beta$-tracks we restrict the analysis to 10~bar, pressure at which, according to simulations, a fraction 45\% of the source energy is contained (compared to an asymptotic value of 55\% at high pressure). The rapid change in containment with pressure is illustrated in Fig. \ref{fig:sources_G4}-right.

\begin{figure}[h!!!]
    \centering
    \includegraphics[scale=0.52]{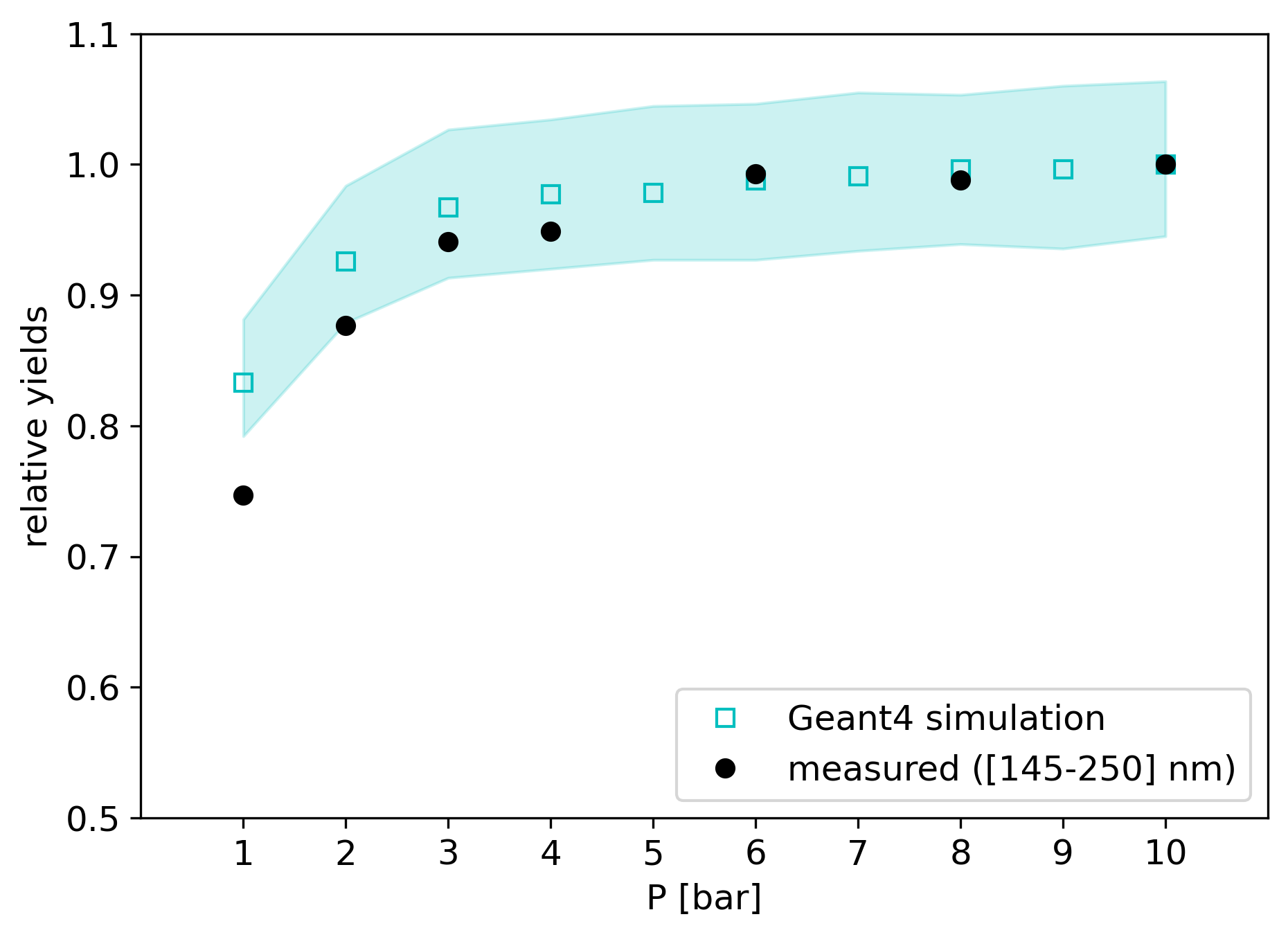}
    \caption{Measured yields in the 145-250~nm band relative to the average of the two highest points and comparison with Geant4 simulations. The band includes the spread on the assumed reflectivity values.
    }
\label{fig:geff}
\end{figure}

A comparison between the experimental measurements of the primary scintillation in the $2^{nd}$ continuum band (obtained at high electric field) and the corresponding Geant4 simulations is shown in Fig. \ref{fig:geff}, normalized to the highest pressure yield. In the following, the average value of the simulated light collection efficiency and its standard deviation is used for correcting the experimental yields.

\section{Results for primary scintillation (S1) \label{S1}}

\subsection{Electric field dependence}

\begin{figure}[h!!!]
    \centering
    \includegraphics[scale=0.6]{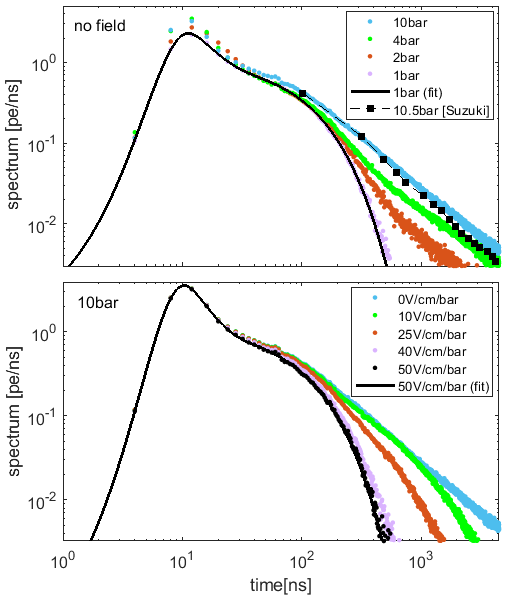}
    \caption{Average time spectra for the primary scintillation (S1) signal as detected with the unfiltered PMs (range 145-600~ns), given in photoelectrons per ns. Top: dependence with pressure for no-field conditions, with data from \cite{suzu2} given as squares. Bottom: dependence with pressure-reduced field at 10~bar. A fit to eq. \ref{ExpFunc} (including a fast and slow component) is given by the continuous line, in recombination-free conditions.}
\label{fig:RecoVst}
\end{figure}

Primary scintillation for $\alpha$-tracks at varying pressure and electric field has been studied thoroughly, for instance in the works of Suzuki, Saito, Mimura, and Bolotnikov and Ramsey ~\cite{suzu1, suzu2, Japa1, Japa2, Ramsey}. Given the high ionization-density, charge recombination becomes severe at high pressure, with each neutralized ion resulting in the emission of an additional photon in correlation. At a shorter time scale, the scintillation process is dominated by the singlet and triplet scintillation precursors, with exponential time constants $\tau_f = 4.5$~ns, $\tau_s=100$~ns (e.g., \cite{XeSim} and references therein). However, as the time exceeds few 100's of ns, a harder power law component can be anticipated following the arguments in \cite{suzu2}, roughly as $dN/dt \sim t^{-1/2}$. The effect, resulting from the charge recombination process, can be seen in Fig. \ref{fig:RecoVst}-top, where the time spectrum in the entire observed range of 145-600~nm is shown for different pressures and no electric field. In Fig. \ref{fig:RecoVst}-bottom the electric-field dependence for 10~bar is also shown. For comparison, measurements from \cite{suzu2} have been overlaid. Despite the longer time scale used in that work (up to $30~\mu$s), both measurements indicate that the amount of recombination light (charge) above the first $5~\mu$s is well below 5\%, even in the harshest conditions considered here (10~bar, no electric field). The figure also highlights how that the prompt component increases with pressure, as discussed in next section.

\begin{figure}[h!!!]
    \centering
    \includegraphics[scale=0.50]{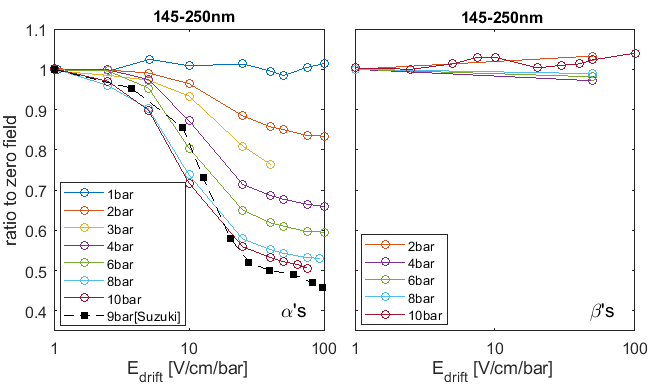}
    \includegraphics[scale=0.5]{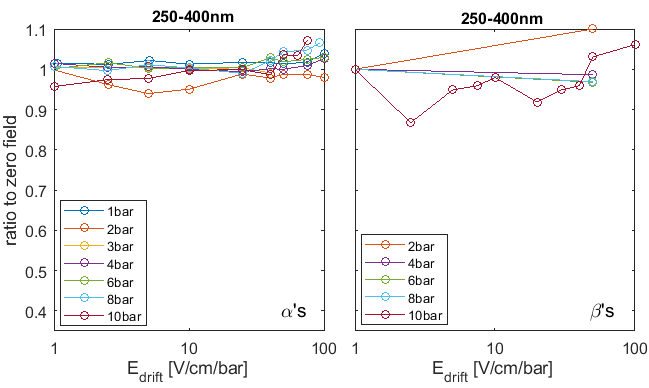}
    \caption{Primary scintillation as a function of electric field, relative to zero-field conditions. Top-left: $\alpha$-particles from an $^{241}$Am source, observed in the band 145-250~nm (data from \cite{suzu1} is shown as black squares). Top-right: $\beta$-particles from a $^{90}$Sr source, band 145-250~nm. Bottom-left: $\alpha$-particles from an $^{241}$Am source, band 250-400~nm. Bottom-right: $\beta$-particles from a $^{90}$Sr source, band 250-400~nm. For the $\beta$ source and pressures below 10~bar, only control measurements at 0 and 50~V/cm/bar were performed.}
\label{fig:RecoBoth}
\end{figure}

Figure \ref{fig:RecoBoth} shows the $2^{nd}$ continuum yields (145-250~nm) for $\alpha$ (top-left) and $\beta$ (top-right) particles obtained on a $5~\mu$s window and normalized to zero-field conditions. For the case of $\alpha$'s, data at 9~bar from \cite{suzu1} has been overlaid (squares). It has to be mentioned that the corresponding figure as presented in that work seems to indicate that charge recombination is independent from pressure above 3~bar, contrary to our observations. When inspected more closely, it appears that only high pressure measurements extend down to zero field, so the normalization for lower pressures seems arbitrary and thus misleading. Under the assumption that $W_{sc}$ is independent from pressure up to 10~bar in the absence of recombination (i.e., at high fields), our observations are more in line with those of Saito \cite{Japa2}, who shows that $W_{sc}(E=0)$ becomes gradually smaller at high pressure in the range 1-10~bar, as observed here (see next section). Figure \ref{fig:RecoBoth}-top/right shows, on the other hand, that charge recombination from $\beta$ particles is negligible even at 10~bar, a pressure sufficient to contain 0.5~MeV tracks (e.g., \cite{Accu}). Control measurements at 0 and 50~V/cm/bar have been performed at lower pressures, showing compatible values. These measurements seem sufficient to convincingly exclude any sizeable form of charge recombination for the case of $\beta$-particles up to 0.5~MeV, in line with the observations for $x$-rays performed also at 10~bar in \cite{Balan}. Higher energy (1-2~MeV) electrons, stemming for instance from $\gamma$-conversions, will certainly undergo even smaller recombination effects, as an increasingly large fraction of the track pertains to the minimum ionizing regime. This result is of interest to contemporary high pressure TPCs (e.g. \cite{NEXT}); its validity, however, is restricted to pure xenon and is not generally applicable in the presence of additives, as the diffusion of the ionization cloud can be reduced by up to one order of magnitude in those conditions \cite{XeHe, XeCO2}.

Also interestingly, the bottom rows in Fig. \ref{fig:RecoBoth} show the same measurements for the scintillation in the 250-400~nm band, that is classically attributed to the $3^{rd}$ continuum and hence involving Xe$_n^{+,*}$, Xe$_n^{++}$ precursor states (e.g. \cite{Wiese}, \cite{Langlorf}, \cite{RusosThird}). Despite the higher ionization state and reactivity compared to single-ionized ions, it is interesting to see that they do not participate of the charge recombination process. Similarly, the time constant for the emission is in the range 8.3-8.5~ns with independence of the applied field, these values being similar to earlier measurements performed in \cite{Millet} (8.2$\pm$0.5~ns). A control measurement on the same band for argon at 10~bar yielded 4.8~ns, also in agreement with expectations (e.g., \cite{Edgar}). Similar conclusions can be extracted for $\beta$-particles: at 10~bar, where 0.5~MeV electron tracks are well contained in our mini-TPC, no systematic trend can be observed, within the somewhat large errors stemming from the lower scintillation in this case. The scintillation in the range 400-600~nm displays identical behaviour and is not shown.

\subsection{Pressure dependence}

Pressure variations cause changes in the scintillation mechanisms as well as in the light collection efficiency, the latter due to changes in the size of the particle trails. For $\alpha$-particles, the primary scintillation yields obtained at high fields in the range [145-250]~nm decrease by up to 25\% at 1~bar relative to 10~bar, in approximate agreement with the decrease of the light collection efficiency obtained in simulation (Fig. \ref{fig:collection_efficiency_a}). The fact that the $2^{nd}$ continuum yields at high fields are dominated by light collection effects, being approximately flat at the production point, is consistent with earlier observations (for instance \cite{Japa2}). 

The absolute yield, on the other hand, can be obtained when considering the measured number of photoelectrons $Y_m$, the weighted quantum efficiency $<QE>_\lambda$, the light collection efficiency ($\mathcal{G}$) and the transparency ($\mathcal{T}$) of the wires (93.6\% for the gating grid, 96.8\% for the cathodes and 99\% for the anode). In particular, $W_{sc}$ can then be found as:
\beq
W_{sc} =  \frac{\mathcal{G} \cdot \mathcal{T} \cdot <QE>_\lambda}{Y_{m}[145-600]- g \cdot Y_{m}[250-600]} \cdot \varepsilon_\alpha \label{Wsc_eq}
\eeq
where $\varepsilon_\alpha$ indicates the energy of the $\alpha$ particle and $g$ is a factor of order one that corrects for the marginally different geometrical efficiencies of filtered and unfiltered PMs.
Using the wavelength response of the different system components in Fig. \ref{lambda_dep}, the average quantum efficiency in the $2^{nd}$ continuum range can be estimated to be ${<QE>_{[145-250]}=0.137}$, while for the third continuum range we obtain ${<QE>_{[250-400]}=0.182}$ and ${<QE>_{[250-600]}=0.189}$. Prior to this work, the third continuum spectra has been studied at around 1~bar in \cite{Millet} ($\alpha$ excitation) and \cite{Robert} (x-ray flash) showing good mutual agreement. The pressure-dependence has been measured in \cite{Henck} ($\alpha$ excitation) up to 4~bar, showing little dependence except for an overall blue-shift of 50~nm compared to the above results. We thus assume the $3^{rd}$ continuum spectrum to be pressure-independent in our conditions. For the $2^{nd}$ continuum, independence of shape above 100's of mbar is well established (e.g., \cite{photoex1}).

\begin{figure}[h!!!]
    \centering
    \includegraphics[scale=0.55]{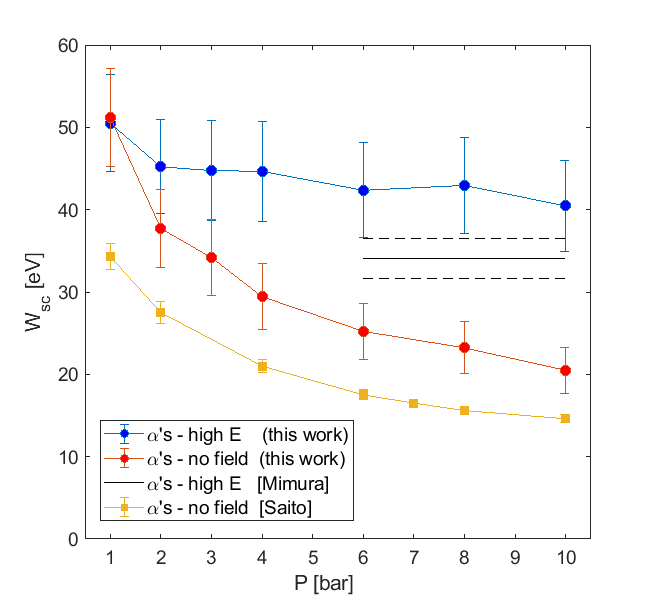}
    \caption{Average energy to produce a scintillation photon in the $2^{nd}$ continuum band ($W_{sc}$) after PM calibration, quantum efficiency and geometrical corrections (blue and red circles). A comparison with (band-unresolved) data is shown both at zero field in \cite{Japa1} (squares) and high field in \cite{Japa2} (average value shown as continuous line and error band as dashed).}
\label{fig:Wsc}
\end{figure}

$W_{sc}$-values estimated from expression \ref{Wsc_eq} are shown in Fig. \ref{fig:Wsc} for the 145-250 nm range (red and blue circles). For comparison, data from \cite{Japa1} at zero-field are shown as squares and the average value obtained at high field in \cite{Japa2} as a continuous line with its error (dashed). Since the scintillation measurements in previous works were not band-resolved, the systematic up-shift observed here might be naturally attributed to the effect of $3^{rd}$ emission, that usually takes place in a more favourable detection band.

\begin{figure}[h!!!]
    \centering
    \includegraphics[scale=0.65]{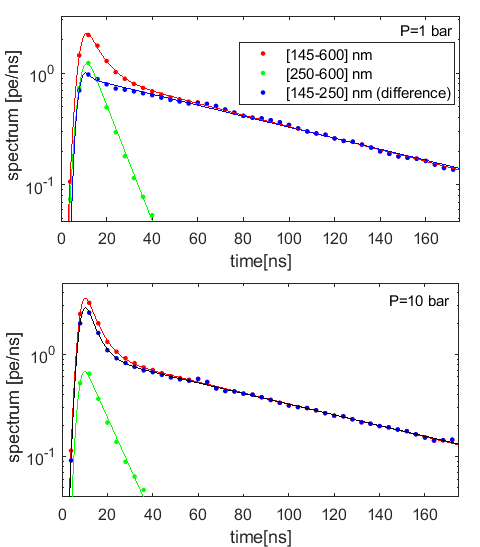}
    \caption{Average time spectra for the primary scintillation (S1) in high electric field conditions (above 50~V/cm/bar, i.e., no charge recombination). The directly measured spectra (red for [145-600]~nm and green for [250-600]~nm) have been subtracted to obtain the uncontaminated $2^{nd}$ continuum signal (blue). Top: 1~bar. Bottom: 10~bar. Fits to eq. \ref{ExpFunc} have been included with continuous lines.}
\label{fig:Pdependence}
\end{figure}

One outstanding feature, characteristic of the scintillation process in this pressure range, is that the scintillation mechanisms responsible for the $2^{nd}$ continuum (determining the populations of the singlet and triplet components) are expected to saturate slightly above 2~bar (see for instance \cite{photoex5}). The high quality of the single-photon calibration and time-alignment of the waveforms allows to reliably subtract the time spectra, so the phenomenon can be addressed more clearly. According to Fig. \ref{fig:Pdependence}-up, in our $2^{nd}$ continuum measurements (blue dots) the singlet component is not even discernible at around 1~bar, however at 10~bar it is already fully formed (Fig. \ref{fig:Pdependence}-bottom). Indeed, the fast component observed over the entire range of [145-600]~nm (red dots) is dominated at 1~bar by the $3^{rd}$ continuum emission (green dots), in analogous way to the observations recently made in argon \cite{Edgar}. So while the time spectrum looks qualitatively independent and relatively dull as a function of pressure, the fast component peak changes from being $3^{rd}$ continuum-dominated to being $2^{nd}$ continuum-dominated, and likewise its spectral content changes too.

The dynamics of the $3^{rd}$ continuum can be studied by analysing its scintillation yield relative to the $2^{nd}$ continuum one, so as to cancel most of the effect from the uncertainty in the geometrical correction factor, that dominates the absolute yields. The fraction of $3^{rd}$ continuum emission is reduced by about a factor $\times 2.5$ when increasing pressure from 1 to 10 bar, starting from a seemingly modest 10\% value and ending at 4\%. However, when considering its effect on the fast component (for instance considering a 20~ns time window around the fast peak) its relative contribution goes from 60\% at 1~bar to 20\% at 10~bar. At high pressures, where the event containment is highest, the ratio has been derived for $\beta$-electrons too (pink point). Yields from $\beta$-electrons emerging from surface deposits are complicated to interpret, as they unavoidably interact with walls, potentially leading to spurious scintillation. Thus, the $2^{nd}/3^{rd}$ continuum ratio has been obtained in this case as an average value with an uncertainty coming from its spread, obtained for different cuts in the number of photoelectrons detected. The approximately 30\% spread observed, compared to up to a factor $\times 20$ obtained in simulation if assuming that scintillation from the walls dominates the $3^{rd}$ continuum signal, suggests that most of the observed scintillation has its origin in the gas.

\begin{figure}[h!!!]
    \centering
    \includegraphics[scale=0.50]{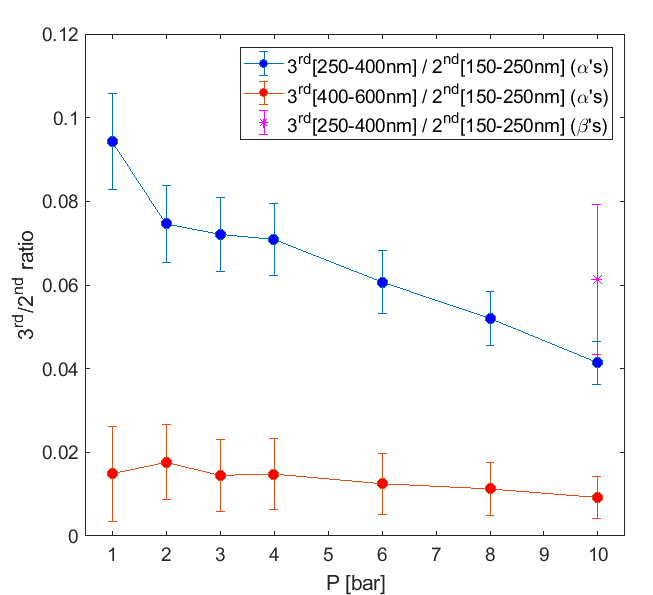}
    \caption{Ratio of $3^{rd}$ to $2^{nd}$ continuum signals for $\alpha$-excitation, considering the 250-400~nm and 400-600~nm bands (blue and red, respectively). The pink point represents the ratio for $\beta$-electrons as obtained at 10~bar, where $\beta$ events are contained up to around 0.5~MeV.}
\label{fig:3rdTo2nd}
\end{figure}

For a complete study, time constants have been obtained from fits to eq. \ref{ExpFunc} and are compiled in table \ref{table}. For the $3^{rd}$ continuum emission (in the range 250-400~nm) they are found to vary with pressure in the range 8.23-8.75~ns (being largely field-independent). These time constants as well as the light yields were found to vary by less than 5\% when the gas purity was purposely brought to conditions where the triplet time constant was only 34~ns, in a dedicated run at 2~bar. Additionally, for the highest pressure (at which the fast peak observed in the unfiltered PMs is expected to be dominated by the singlet component of the $2^{nd}$ continuum) a value for $\tau_f = 5.2\pm1.0$~ns was obtained, compatible with the singlet time constant. Values for the triplet time constants are affected by impurities, so we select the highest value as an estimate, corresponding to injection of fresh gas at 2~bar ($\tau_s =98.5\pm0.5$~ns).

\begin{table}[h!!!]
\begin{tabular}{| c | c | c | c | c |}
\hline
$\tau_{f}|_{3^{rd}}$[ns]   & $~8.30\pm0.07$~ & $~8.16\pm0.10~$ & $~8.63\pm0.07~$ & $~8.50\pm0.25~$ \\
\hline
P [bar]         & 1 & 2 & 4 & 10 \\
\hline
\end{tabular}
\caption{Time constants obtained in this work for the $3^{rd}$ continuum band at different pressures (largely independent from the pressure-reduced field in the range 0-100~V/cm/bar). The value for the fast component of the $2^{nd}$ continuum, obtained from a fit to the unfiltered PM in high pressure conditions (to minimize contamination from $3^{rd}$ continuum emission) was $\tau_{f}|_{2^{nd}} = 5.2\pm 1.0$~ns while the highest value achieved for the slow component (at $E>50$~V/cm/bar) was $\tau_{s}|_{2^{nd}} =98.5\pm 0.5$~ns. \label{table}}
\end{table}

\begin{figure}[h!!!]
    \centering
    \includegraphics[scale=0.31]{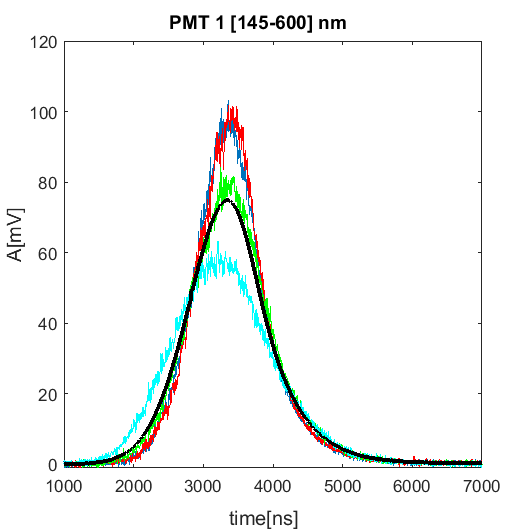}
    \includegraphics[scale=0.31]{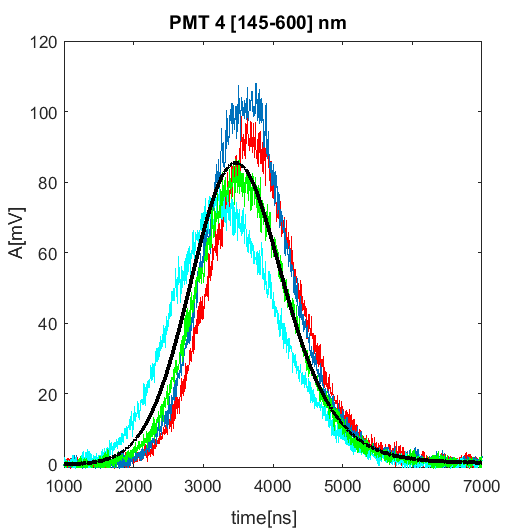}
    
    \includegraphics[scale=0.31]{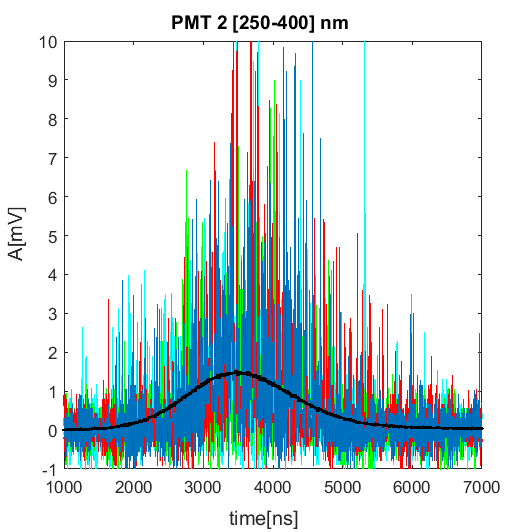}
    \includegraphics[scale=0.31]{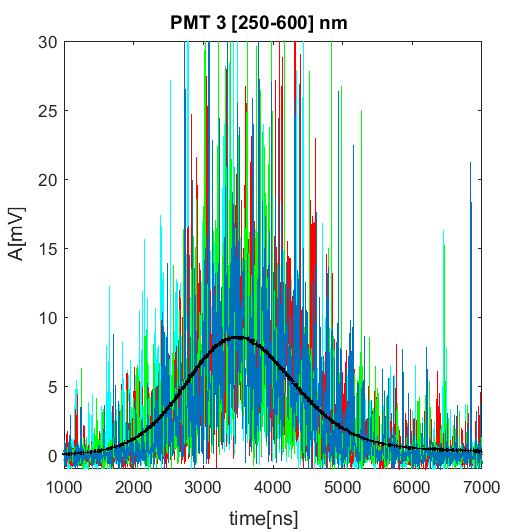}
    
    \caption{Several waveforms recorded in the 145-600~nm range (upper plots), 250-400~nm (bottom-left) and 250-600~nm (bottom-right). The average waveform obtained for 10000~evts is shown in black (thick line). The events were taken at 10~bar, with 4.0~kV applied at the anode and 50~V/cm/bar in the drift region. (Amplitudes are not directly comparable as the PM bias is different in each case).}
\label{SelectionS2}
\end{figure}

\section{Results for secondary scintillation (S2) \label{S2}}

Secondary scintillation was obtained following an analysis procedure similar to the one used in the reconstruction of the primary scintillation. Aiming at a direct analysis, we concentrated on the pressure range 3-10~bar to avoid geometrical effects that complicate the reconstruction of the time profile due to the extension of the ionization trail. In the chosen pressure regime the events are well-described as diffusion-dominated Gaussian functions.
In all cases, a drift field of 50~V/cm/bar was employed in order to avoid sizeable charge-recombination effects. Importantly, in order to overcome PM saturation during the scan, the PM bias voltage was varied so as to always being below a predefined maximum signal. The lack of saturation was confirmed offline by seeking linearity in the charge vs. amplitude plots (the former being more prone to saturation). An example of an average event in the different bands is given in Fig. \ref{SelectionS2}.

\begin{figure}[h!!!]
    \centering
    \includegraphics[scale=0.5]{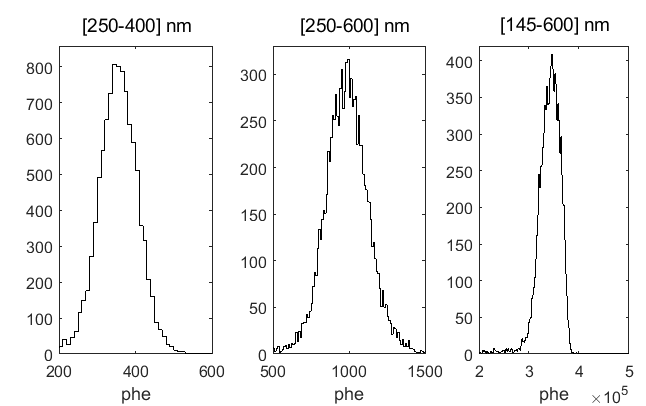}
    \caption{Example of photoelectron distributions obtained under $\alpha$-tracks at a pressure of 10~bar, with 4.0~kV applied at the anode and 50~V/vm/bar in the drift region.}
\label{QS2}
\end{figure}

For each event, the signal charge (integral of the waveform) was converted to photoelectrons (phe) based on LED calibrations and histogrammed. Soft cuts in low energy events were applied to avoid outliers, and the average value of the histograms was used as an estimate of the yields (Fig. \ref{QS2}).

\begin{figure}[h!!!]
    \centering
    \includegraphics[scale=0.5]{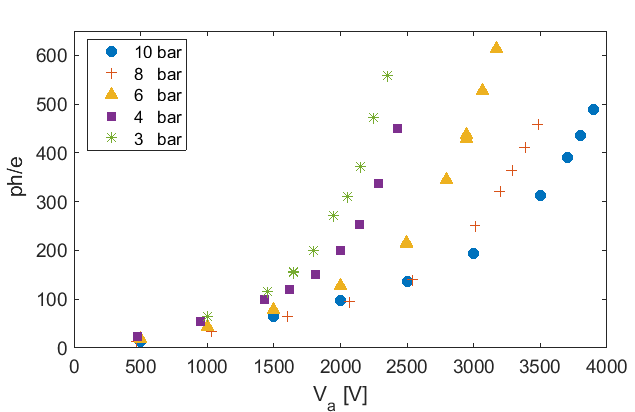}
    \includegraphics[scale=0.5]{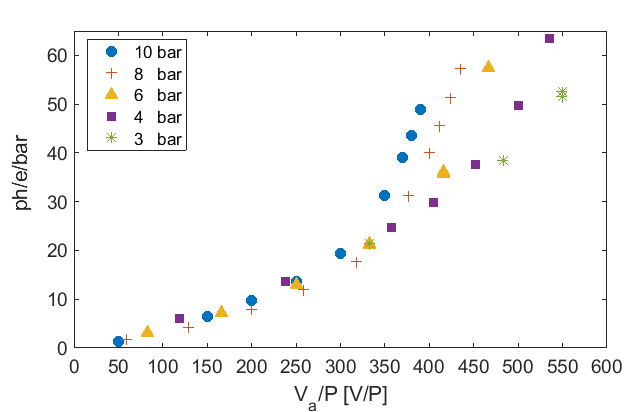}
    \caption{Top: absolutely normalized secondary scintillation yields as a function of pressure, measured by the unfiltered PM (145-600 nm). Bottom: scintillation yields in a pressure-reduced representation. The drift field is 50~V/cm/bar in all cases and the overall systematic error is estimated to be 25\%.}
\label{yields}
\end{figure}

By applying the Geant4 corrections with the same parameters as for the primary scintillation signals, and given the good agreement obtained with earlier measurements of the primary scintillation, one can expect that the reported secondary scintillation yields are within less than a $25\%$ error. This uncertainty can be decomposed as coming from two sources, added quadratically: 10-13\% represents the measurement error (dominated by the single-photon calibration and the geometrical efficiency) and 20\% is the maximum deviation with respect to earlier $W_{sc}$ values, added in quadrature. Fig. \ref{yields} shows the absolutely normalized secondary scintillation yields obtained in this way (top) an after applying $P$-scalings (bottom).

The measurements show clearly the presence of a proportional EL-region together with an exponential one, hinting at the onset of multiplication. As expected, the linear region shows $yield/P$ vs. $E/P$ scalings while the exponential one does not. Indeed, the multiplication process follows the analogous scaling for the yield of electrons per cm, however this latter magnitude enters in an exponent due to the avalanche-nature of the process. As a result, for the same pressure-reduced voltage it is easier to obtain gain at higher pressure, as Fig. \ref{yields} (bottom) illustrates. More important, we can not reproduce here the evidence pointing to the violation of $P$-scalings reported earlier in \cite{Pviolation}. In that work, an increase of the pressure-reduced yield was observed as a function of pressure, at fixed values of the pressure-reduced electric fields. Our measurements, on the other hand, favour a canonical explanation without the need to invoke new phenomena (for instance, increased probability of excitations of ground-state dimers as the pressure increases).

\begin{figure}[h!!!]
    \centering
    \includegraphics[scale=0.5]{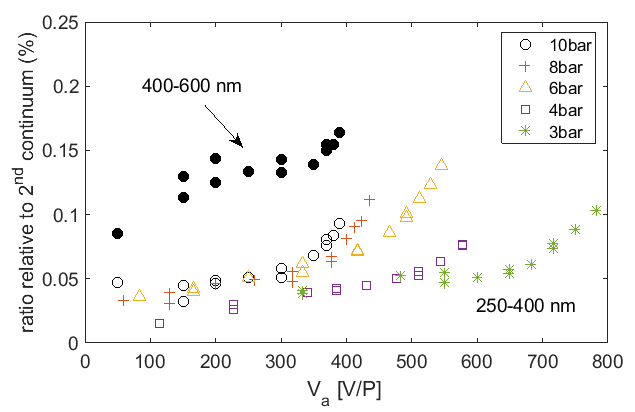}
    \caption{Ratio of detected light yields in the UV region [250-400]~nm for different pressures. For illustration, the (stronger) emission in the visible [400-600]~nm band is shown too.}
\label{ratios_vs_Va}
\end{figure}

Finally, Fig. \ref{ratios_vs_Va} shows the scintillation in the 250-400~nm and 400-600~nm ranges relative to the yield in the range 145-250~nm (expectedly, from the $2^{nd}$ continuum). The emission in the UV region (250-400~nm) is nearly flat at 0.05\% level and increases exponentially with the onset of multiplication, suggesting that the contribution of highly-lying states starts to become more important relative to $2^{nd}$ continuum emission. The visible emission in the range 400-600~nm (plot just at 10~bar, for clarity) shows a much lower dependence with the onset of multiplication, suggesting a dominant contribution from a process independent from excimer production. A natural candidate is neutral bremsstrahlung, since the 0.15\% levels reported here on the range 400-600~nm are compatible with the 0.1-1\% levels expected for the effect when considering high EL fields \cite{nBrXe2}. Also, as expected, yields are higher than in the UV region. Irrespective from the potential of further studies along this direction, it is interesting to note that $\alpha$ tracks can be cleanly reconstructed in either of the bands, not requiring the detection of the $2^{nd}$ continuum band (e.g., Fig. \ref{QS2}).

\section{Conclusions}

Simultaneous time, band and particle -resolved measurements of primary (S1) and secondary (S2) scintillation have been performed in pure xenon through a mini time projection chamber (TPC), its anode equipped with a multi-wire readout. The systematic measurements performed as a function of pressure and electric field allow a thorough characterization of subdominant phenomena in xenon. In particular, we have been able to establish the relative S1-yields of 3$^{rd}$ to 2$^{nd}$ continua emission for $\alpha$-particles ($^{241}$Am) to be in the range 10-4\% in the pressure range 1-10~bar. At 10 bar, where energy containment reaches 80\% of its maximum value, $\beta's$ from a $^{90}$Sr-source produce a ratio of $6 \pm 2$\%, compatible with the value observed for $\alpha$'s. Similarly to the recent observations in argon \cite{Edgar}, for the lowest pressures studied (around 1~bar in case of xenon) the prompt scintillation component is dominated by this 3$^{rd}$-continuum emission thus creating, in the absence of spectral information, the illusion of a pressure-independent time profile. This 3$^{rd}$ continuum emission has a time constant in the range 8.3-8.5~ns, compatible with previous observations, being the yields and time constants largely field-independent in the measured pressure range, up to pressure-reduced fields of 100~V/cm/bar. On the other hand, the $2^{nd}$ continuum yields for $\alpha$-particles are compatible with earlier values although somewhat larger by 20\% ($W_{sc} = 43 \pm 5$ eV in the range 1-10~bar). The well-studied light recombination characteristics are reproduced at the same level. Although part of the difference can be explained through the previously unaccounted $3^{rd}$ continuum emission, some tension remains. In the case of $\beta$-particles, no hint of recombination light was observed even at 10~bar and in the absence of electric field, as anticipated.

Last, measurements of the secondary scintillation component exhibit a good pressure-scaling in the 3-10~bar range, against earlier evidence \cite{Pviolation}. Our results favour a canonical interpretation without the need to invoke the dynamics of neutral Xe$^*$ dimers in the gas. The band-resolved measurements performed in the range 400-600~nm and 250-400~nm show the presence of additional secondary scintillation at the level of 0.05-0.2\%, qualitatively compatible with the intensity and spectral characteristics of neutral bremsstrahlung. Within these small values, a stronger participation of excited states in the 250-400~nm range is observed, an effect particularly noticeable when the proportional regime is abandoned. 
\ack

This research is sponsored by the IGFAE-IGNITE initiative, and by the Spanish Ministry of Economy and Competitiveness (MINECO) through the grant FPA2017-92505-EXP. DGD is supported by the Ram\'on y Cajal program (Spain) under contract number RYC-2015-18820. The authors want to acknowledge NEXT and RD51 members for insightful discussions, specially J. Santos, C. Monteiro, C. Henriques (Univ. Coimbra) and Justo Mart\'in-Albo (IFIC) for sharing their gigantic knowledge on the aspects of xenon scintillation.

\end{document}